\documentstyle[aps,prb,eqsecnum,epsfig]{revtex}
\begin{document}
\author{Mikio {\sc Eto}}
\address{Faculty of Science and Technology, Keio University, \\
3-14-1 Hiyoshi, Kohoku-ku, Yokohama 223-8522, Japan}
\title{Electronic States and Transport Phenomena in Quantum Dot
Systems}
\date{September 10, 2000}
\maketitle
\begin{abstract}
Electronic states and transport phenomena in semiconductor
quantum dots are studied theoretically.
Taking account of the electron-electron Coulomb interaction
by the exact diagonalization method,
the ground state and low-lying excited states are
calculated as functions of magnetic field.
Using the obtained many-body states, we discuss
the temperature dependence of the conductance
peaks in the Coulomb oscillation.
In the Coulomb blockade region, elastic and inelastic
cotunneling currents are evaluated under finite bias
voltages.
The cotunneling conductance is markedly enhanced by the Kondo effect.
In coupled quantum dots, molecular orbitals
and electronic correlation influence the transport
properties. \vspace*{0.5cm} 
\end{abstract}

\section{Introduction}

In disk-shaped quantum dots fabricated on semiconductors,
discrete energy levels show an atomic-like shell
structure.\cite{Tarucha}
These energy levels are filled consecutively
with increasing the number of electrons, as shown in the periodic
table of atoms.
By coupling quantum dots, molecular orbitals between the dots
have been observed.\cite{Oosterkamp,Blick}
In these artificial atoms and molecules, we can
control several parameters, {\it e.g.},
number of electrons, strength of Coulomb interaction,
coupling to external leads, using gate and bias voltages,
applying magnetic fields and so forth, and even realize situations
which are not possible in conventional solid-state systems.
The fundamental study of these systems is
exploring new and rich physics.

In this paper we review our theoretical work on electronic states
in quantum dots and discuss various transport phenomena.
We examine many-body states for a few electrons confined in the dots.
In ``vertical'' quantum dots, the strength of the electron-electron
Coulomb interaction is comparable to the discrete level
spacing.\cite{Tarucha}
To calculate the electronic states, we adopt the exact
diagonalization method which is the best way to take account of
the Coulomb interaction.\cite{qd1,qd2,qd3,spinblock,Daniela,me1,me2}
When a magnetic field $H$ is applied perpendicularly to the dots,
the wavefunctions are shrunk and hence the interaction
becomes more effective. As a result, the ground state changes to a
strongly correlated state at some values of $H$.
The transitions of the ground state have been observed
experimentally, in agreement with our calculations.\cite{Leo}

The electronic states in the dots influence the transport properties.
The conductance peaks of the Coulomb oscillation are examined
using the obtained many-body states.
The peak height exhibits an unusual
temperature dependence when more than one level contributes to the
current. Some peaks are suppressed by the
``spin blockade,''\cite{spinblock}
in which the transition between the ground
states of $N$ and $N+1$ electrons is forbidden by the spin selection
rule.\cite{me3}

In the Coulomb blockade regions, 
quantum-mechanical higher-order processes play an important
role in the transport.
They are called ``cotunneling'' and cause serious
current leakage when the quantum dots are applied to single electron devices.
There are two kinds of cotunneling processes, elastic and inelastic
processes.\cite{Yuli1} At low bias voltages, only the elastic
process is possible. When the bias voltage
exceeds the excitation energy in a quantum dot, both the elastic and
inelastic processes contribute to the current.\cite{Funabas1,Funabas2}
When a localized spin exists in a dot, the Kondo effect
markedly enhances the cotunneling current. The Kondo effect
in quantum dots is being studied
experimentally.\cite{Kondo1,Kondo2,Kondo3,Kondo4,Kondo5}

In coupled quantum dots,
molecular orbitals are formed when the dot-dot tunnel
coupling is sufficiently strong.\cite{Kawamura}
These systems have attracted recent interest with respect to
their application to quantum computing.\cite{Loss1,Loss2}
We examine the transport properties of the dots connected in series,
which reflect both the molecular orbitals and electron-electron
interaction.\cite{me4}

The organization of this paper is as follows.
In the next section (\S 2), the electronic states confined
in a quantum dot are examined. Not only the ground state but also
low-lying excited states are calculated. 
Sections 3 and 4 are devoted to an explanation of
the transport properties:
the conductance peaks of the Coulomb oscillation and
cotunneling phenomena in the Coulomb blockade region.
In \S 5, we discuss the electronic states
and transport properties in coupled quantum dots.
The conclusions are given in the last section (\S 6).

\section{Electronic States in Quantum Dots}

We consider $N$ electrons confined in a quantum dot of disk shape,
which is well described by a two-dimensional
harmonic potential.\cite{Tarucha} The electrons interact with
each other by the Coulomb interaction.
A magnetic field $H$ is applied perpendicularly to
the potential plane. The Hamiltonian reads
\begin{eqnarray}
{\cal H} & = & {\cal H}_1 + {\cal H}_2, \\
{\cal H}_1 & = &
\sum_{i=1}^N \left[ \frac{1}{2m^*}
\left({\bf p}_i+\frac{e}{c} {\bf A}({\bf r}_i) \right)^2 
+\frac{1}{2} m^*\omega_0^2 {\bf r}_i^2 \right], \nonumber \\
{\cal H}_2 & = &
\sum_{i<j} \frac{e^2}{\varepsilon|{\bf r}_i-{\bf r}_j|}, \nonumber
\end{eqnarray}
where $m^*$ is the effective mass of electrons and $\varepsilon$ is the
dielectric constant ($m^*=0.067m$, $\varepsilon\approx 13$ in GaAs).
${\bf A}({\bf r})$ is the vector potential.
The Zeeman effect is neglected.

\begin{figure}

\epsfig{file=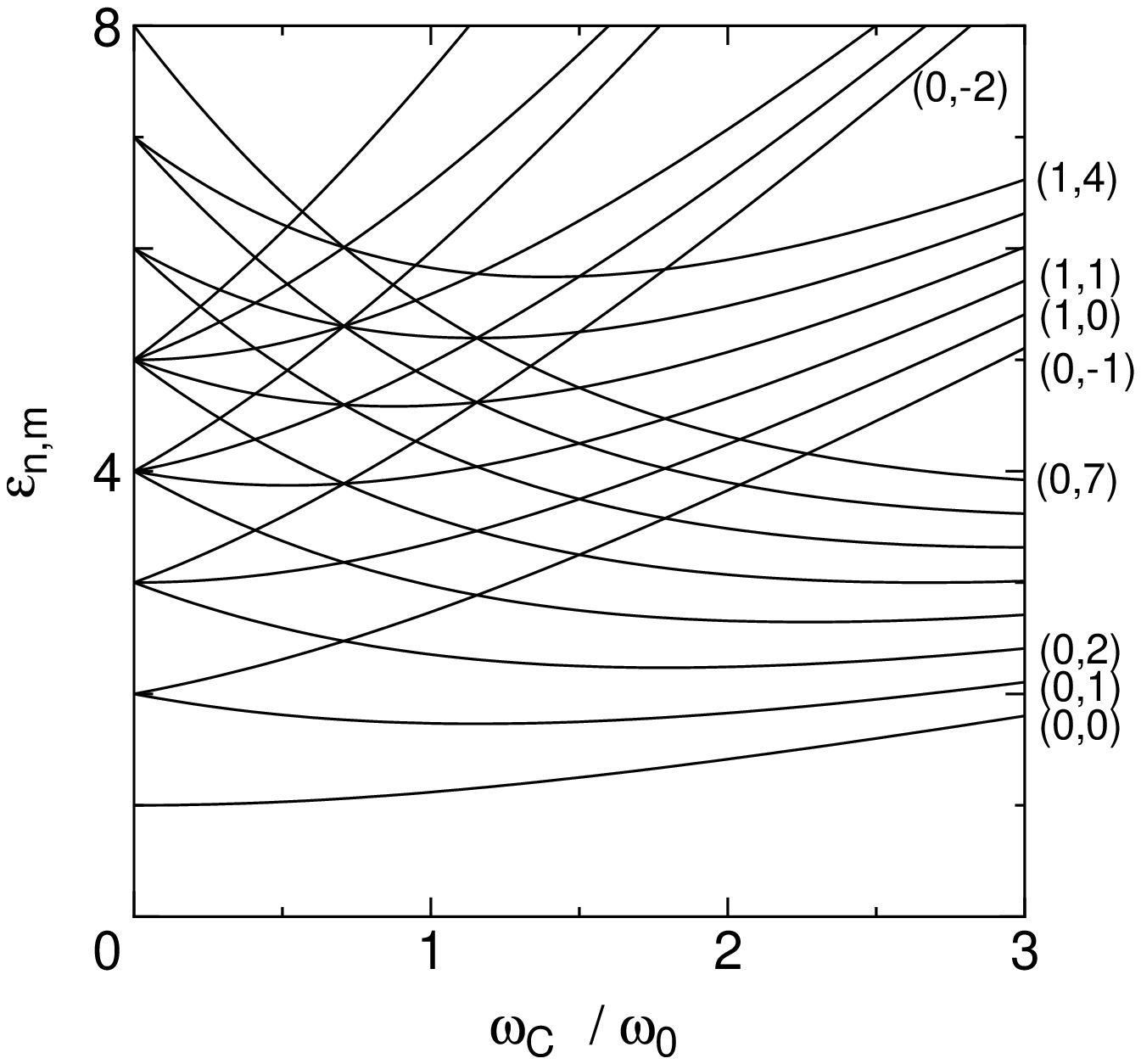,width=8cm}

\caption{
The magnetic field dependence of the one-electron energy levels,
$\varepsilon_{n,m}$, in units of $\hbar\omega_0$.
In the abscissa, $\omega_{\rm c}=eH/m^*c$ is the cyclotron frequency.
The radial and angular momentum quantum numbers, $(n,m)$, are indicated.
\label{fig:fig1}}
\end{figure}

The one-electron part of the Hamiltonian, ${\cal H}_1$, is first
diagonalized to obtain one-electron energy levels. In the absence of
a magnetic field, the energy levels form shells.
The $j$th shell consists of $j$ degenerate levels,
or $2j$ states if the electron spins are counted.
The level spacing is $\hbar\omega_0$, which is $\approx 5$ meV in the
experiments.\cite{Tarucha,Leo}
Under magnetic fields, the one-electron levels are expressed as
\begin{equation}
\varepsilon_{n,m}=\hbar\Omega(H) (2n+|m|+1)
-\frac{1}{2}\hbar\omega_{\rm c}m,
\end{equation}
with
\begin{equation}
\Omega(H)=\sqrt{\omega_0^2+\omega_{\rm c}^2/4},
\end{equation}
where $n$ $(=0,1,2,\cdots)$ and $m$ $(=0,\pm 1,\pm 2,\cdots)$ are
radial and angular momentum quantum numbers,
respectively. $\omega_{\rm c}=eH/m^*c$ is the cyclotron frequency.
The magnetic field dependence of the levels is shown in Fig.~\ref{fig:fig1}.

The strength of the Coulomb interaction, $e^2/\varepsilon l_0$,
where $\displaystyle l_0=\sqrt{\hbar/m^*\omega_0}$ is the dot size,
is comparable to the level spacing.\cite{Tarucha,Leo}
The electron-electron interaction is taken
into account by the exact diagonalization method.\cite{me1}
The Hamiltonian, ${\cal H}$, is diagonalized in a restricted
space of configurations for the occupation of one-electron
levels.\cite{com1} As a result, the electronic states are
represented by linear combinations of many configurations (Slater
determinants). By this method, the correlation effect is
considered beyond the mean field approximation in which
electronic states are described by a single Slater determinant
in the presence of the Hartree and exchange terms.
We obtain the ground state, $\Psi_{N,1}$,
with energy $E_{N,1}$, and low-lying excited states,
$\Psi_{N,i}$, with $E_{N,i}$ ($i=2,3,\cdots$). The addition energies
which are needed to place the $N$th electron on the dot are given by
\begin{equation}
\mu_{N,i}=E_{N,i}-E_{N-1,1},
\end{equation}
where
$\mu_{N,1}$ corresponds to $E_{\rm gate}=\eta eV_{\rm gate}$ at the $N$th
peak of the linear conductance, $V_{\rm gate}$ is the
gate voltage attached to the dot, and $\eta$ is a ratio of the
gate capacitance $C_{\rm gate}$ to the total capacitance
$C_{\rm total}$. $\mu_{N,i}$ ($i\ge 2$)
can also be investigated experimentally by looking at the
differential conductance under finite bias voltages.\cite{Leo}

\begin{figure}
\vspace*{0.3cm}

\noindent
{\large \bf (a)} \hspace{3.5cm} {\large \bf (b)}
\vspace{-0.8cm} \\
\hspace*{-0.5cm}
\begin{minipage}{40mm}
\begin{center}
\epsfig{file=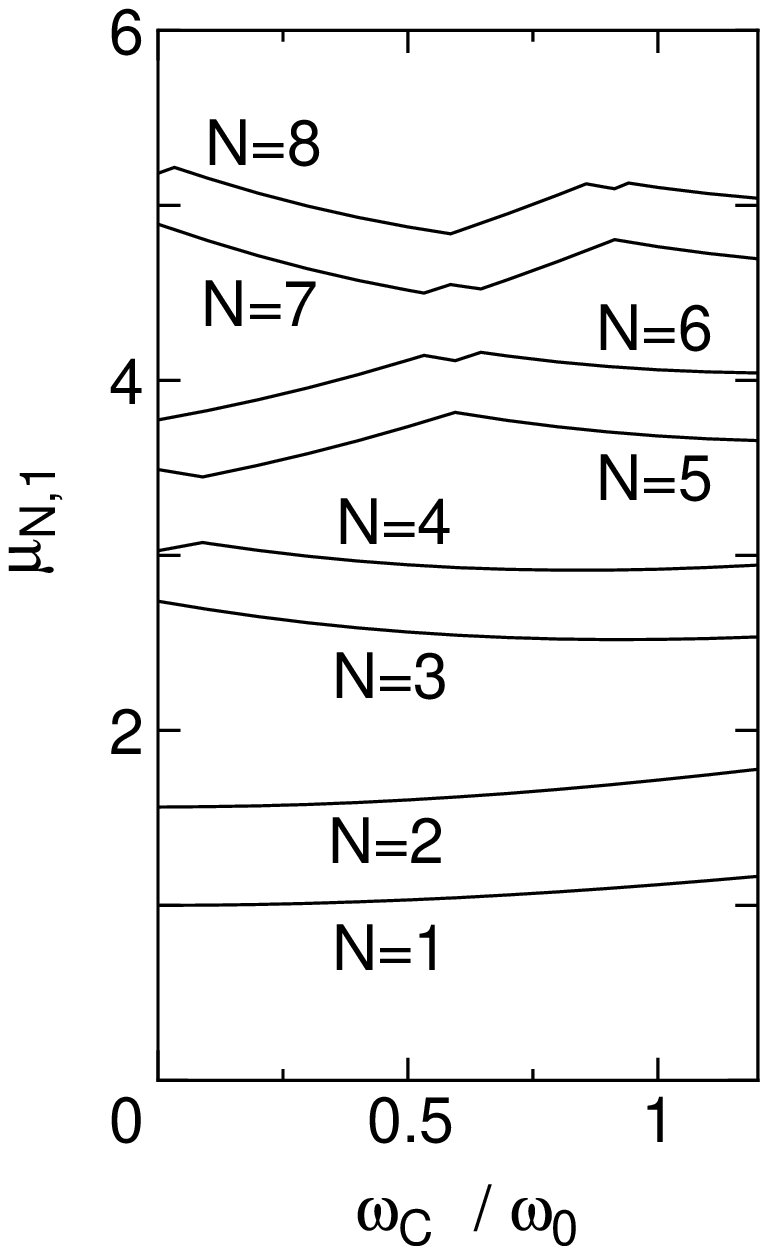,width=5.5cm}
\end{center}
\end{minipage}
\begin{minipage}{40mm}
\begin{center}
\hspace*{0.5cm}
\epsfig{file=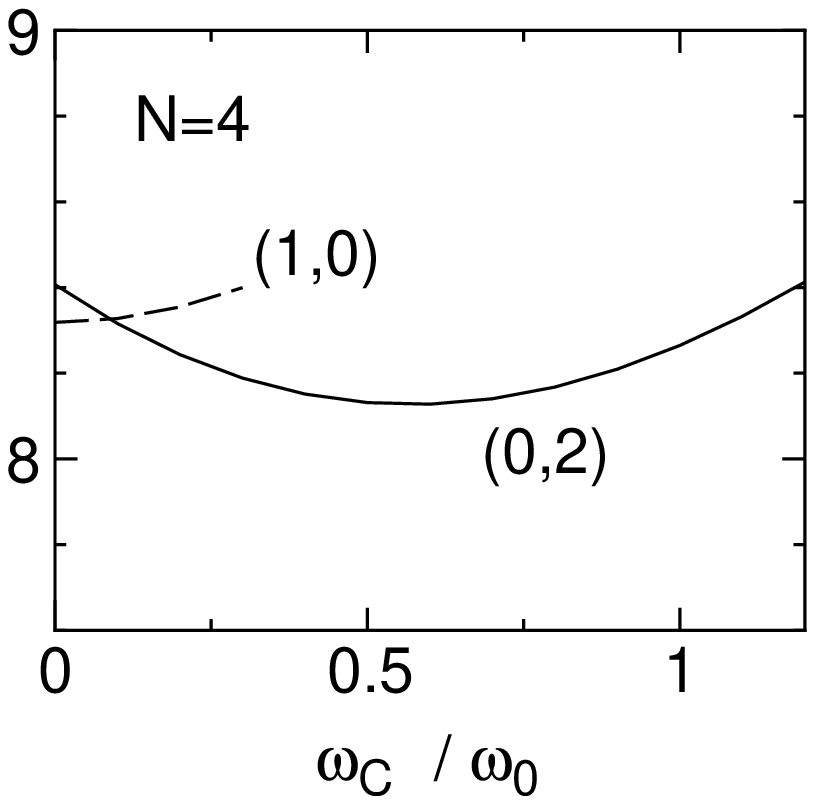,width=4.5cm}
\vspace*{-1.2cm} \\
\hspace*{0.5cm}
\epsfig{file=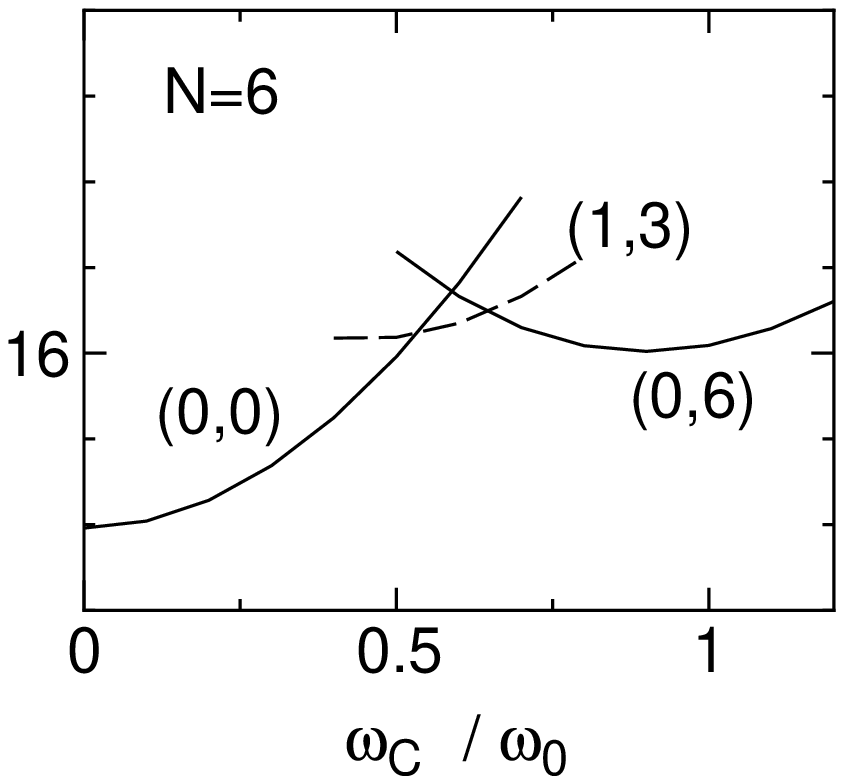,width=4.5cm}
\end{center}
\end{minipage}
\caption{
(a) The magnetic field dependence of the addition energies, $\mu_{N,1}$,
on a quantum dot. The unit of $\mu_{N,1}$ is $\hbar\omega_0$.
In the abscissa, $\omega_{\rm c}=eH/m^*c$ is the cyclotron frequency.
(b) The ground-state energies for $N=4$ and $N=6$ as functions of
magnetic field. The total spin and total angular momentum,
$(S,M)$, are indicated for each state.
[The strength of the Coulomb interaction is
$e^2/\varepsilon l_0=\hbar\omega_0/2$.]
\label{fig:fig2}}
\end{figure}

Figure \ref{fig:fig2}(a) shows peak positions of the Coulomb oscillation
(addition energies, $\mu_{N,1}$), as functions of magnetic field.
When the magnetic field is not too high, the ground state can be
approximated by a single configuration: the one-electron levels shown in
Fig.~\ref{fig:fig1} are filled consecutively.
The magnetic field dependence of $\mu_{N,1}$ is qualitatively understood
by that of the one-electron level occupied by the $N$th electron.
The behavior of $\mu_{1,1}$ and $\mu_{2,1}$ $[\mu_{3,1}$ and $\mu_{4,1}]$
is very similar to that of level $(n,m)=(0,0)$ $[(0,1)]$
in Fig.~\ref{fig:fig1}.
The magnetic field dependence of $\mu_{5,1}$ and $\mu_{6,1}$
indicates that the fifth and sixth electrons occupy the level $(0,-1)$
at $\omega_{\rm c}/\omega_0<0.6$ and $(0,2)$ at $\omega_{\rm c}/\omega_0>0.6$.

Small cusps of $\mu_{4,1}$, $\mu_{5,1}$ ($\mu_{6,1}$, $\mu_{7,1}$)
around $\omega_{\rm c}/\omega_0=0.1$ ($0.6$) are due to the spin-triplet
state for $N=4$ $(6)$ shown in Fig.~\ref{fig:fig2}(b).
The exchange interaction causes high spin states in the vicinity of the
level crossings, between $(0,1)$ and $(0,-1)$ at $\omega_{\rm c}=0$,
and between $(0,-1)$ and $(0,2)$ at $\omega_{\rm c}/\omega_0=1/\sqrt{2}$.
This is the well-known Hund's rule of atoms, which has been confirmed
experimentally.\cite{Tarucha,Tarucha2}

With increasing magnetic field, the level spacings become smaller
(see Fig.~\ref{fig:fig1}) whereas the Coulomb interaction
($\sim e^2/\varepsilon l(H)$) is 
more effective since the wavefunctions are shrunk
as the magnetic length
$\displaystyle l(H)=\sqrt{\hbar/m^*\Omega(H)}$.\cite{me1}
At a critical value of the magnetic field, some electrons start to
occupy higher energy levels to reduce the interaction energy, which results
in a transition of the ground state.
Such correlation-induced transitions are observed when
$\omega_{\rm c}/\omega_0>1.5$.

\begin{figure}

\begin{center}
\epsfig{file=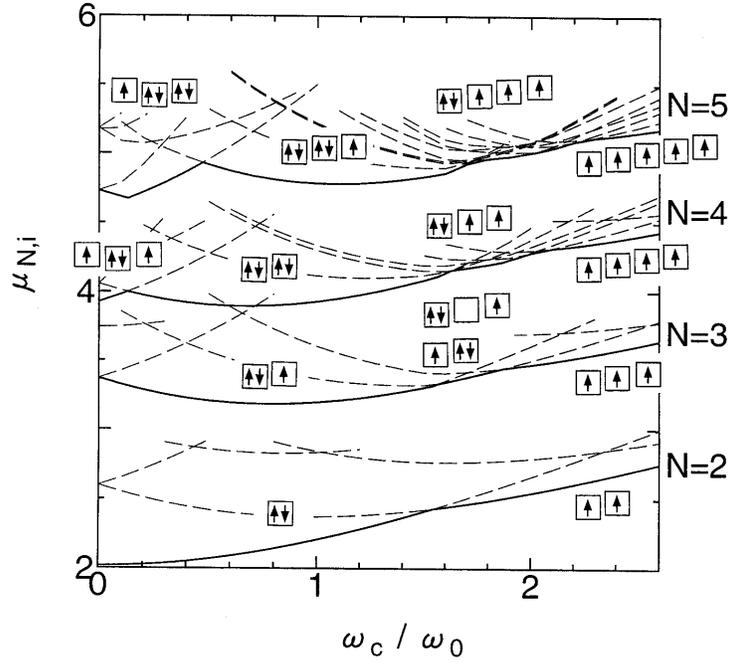,width=10cm}
\end{center}
\caption{
The magnetic field dependence of the addition energies, $\mu_{N,i}$,
for $N=2$ to 5. The addition energies to the ground state,
$\mu_{N,1}$, are represented by solid lines whereas those to excited
states, $\mu_{N,i}$ ($i\ge 2$), are represented by broken lines.
Their unit is $\hbar\omega_0$.
In the abscissa, $\omega_{\rm c}=eH/m^*c$ is the cyclotron frequency. 
For each ground state, the electronic configuration of the largest
amplitude is shown; the lowest square represents the state $(n,m)=(0,0)$.
For squares to the right, angular momentum $m$ increases to 1, 2, 3,
$\cdots$ with $n=0$, and so forth.
For $N=3$, there are two important configurations
in the middle region of the magnetic field.
[The strength of the Coulomb interaction is
$e^2/\varepsilon l_0=\hbar\omega_0$.]
\label{fig:fig3}}
\end{figure}

Figure \ref{fig:fig3} shows the addition energies, $\mu_{N,1}$,
by solid lines
and $\mu_{N,i}$ ($i\ge 2$) by broken lines, to a higher magnetic field
than in Fig.~\ref{fig:fig2}. The transition of the ground state occurs
when a broken line crosses a solid line. For each ground state,
the electronic configuration of the largest amplitude is shown in the
figure.
In the ground state of $N=2$, two electrons occupy level
$(n,m)=(0,0)$ at low magnetic fields. An electron starts to occupy level
$(0,1)$ at $\omega_{\rm c}/\omega_0=1.54$, with parallel spin to the other
electron remaining at level $(0,0)$. As a result, the ground state changes
from a spin singlet ($S=0$) to triplet ($S=1$).\cite{Wagner} For larger $N$,
similar transitions of the ground state occur, one after another,
with increasing magnetic field.
These transitions were observed experimentally.\cite{Leo}
The magnetic field dependence of low-lying excited states as well as
the ground state, shown in Fig.~\ref{fig:fig3},
is in quantitatively good agreement
with the experimental results.\cite{Leo}

In a region of the highest magnetic field in Fig.~\ref{fig:fig3}, 
$N$ electrons are completely spin-polarized (total spin $S=N/2$) and
occupy the lowest $N$ levels. This state is called the maximum density
droplet (MDD). The properties of the MDD and electronic states
under higher magnetic field have been studied experimentally\cite{MDD}
and theoretically.\cite{me2,Natori}

\section{Transport Properties (1) Coulomb Oscillation}

We investigate the transport properties using the obtained many-body
states in the previous section. The coupling between
a dot and two external leads is expressed by the tunneling
Hamiltonian\cite{spinblock,Beenakker,Tanaka}
\begin{equation}
{\cal H}_{\rm T}=\sum_{\alpha={\rm L/R}}\sum_{k,l,\sigma}
\left(
V_{\alpha,l} d_{l,\sigma}^{\dagger} c_{\alpha,k,\sigma}
+ {\rm h.c.}\right),
\end{equation}
where $d_{l,\sigma}^{\dagger}$ and $d_{l,\sigma}$
($c_{\alpha,k,\sigma}^{\dagger}$ and $c_{\alpha,k,\sigma}$)
are the creation
and annihilation operators of state $l$ in the dot
(state $k$ in lead $\alpha$) with spin $\sigma$, respectively.

In this section, the conductance through a quantum dot is calculated
to the lowest order of the tunneling Hamiltonian.\cite{com0}
The transition probability between states $\Psi_{N,i}$ and
$\Psi_{N+1,j}$, with an electron tunneling through the left/right barrier,
is given by
\begin{equation}
\Gamma_{N+1,j; N,i}^{\rm L/R} = \frac{2\pi}{\hbar}
D^{\rm L/R} \left| \sum_{l,\sigma} V_{{\rm L/R},l}
\langle \Psi_{N+1,j} | d_{l,\sigma}^{\dagger} | \Psi_{N,i} \rangle
\right|^2,
\end{equation}
where $D^{\alpha}$ is the density of states in lead $\alpha$.
The linear conductance is written as
\begin{eqnarray}
G=\frac{e^2}{k_{\rm B}T} \sum_{N} \sum_{i,j}
\frac{\Gamma_{N+1,j; N,i}^{\rm L} \Gamma_{N+1,j; N,i}^{\rm R}}
{\Gamma_{N+1,j; N,i}^{\rm L}+\Gamma_{N+1,j; N,i}^{\rm R}}
P_{N,i} \nonumber \\
\times f(E_{N+1,j}-E_{N,i}-E_{\rm gate}-E_{\rm F}),
\end{eqnarray}
where $P_{N,i}$ is the grand-canonical distribution function,
$\displaystyle \exp[-(E_{N,i}-N(E_{\rm gate}+E_{\rm F}))/k_{\rm B}T]/\Xi$,
with the partition function
$\displaystyle \Xi=\sum_{N,i}
\exp[-(E_{N,i}-N(E_{\rm gate}+E_{\rm F}))/k_{\rm B}T]$,
and $f$ is the Fermi distribution function.\cite{Beenakker}

The peaks of the Coulomb oscillation usually become lower and broader
with an increase in temperature $T$.
This is due to the thermal broadening of the electronic distribution
around the Fermi level in the leads.
In experimental results,\cite{Leo2} however,
the heights of the third, fifth, 10th and 17th peaks increase with
temperature, in the absence of magnetic field.
We propose possible mechanisms for this anomalous $T$ dependence of
the peak heights, considering small anisotropy and anharmonicity
of the quantum dot,
\begin{equation}
{\cal H}_{\rm D}^{\prime}/\hbar\omega_0=\lambda_1 (x^2-y^2)/l_0^2
+\lambda_2 (r/l_0)^4.
\end{equation}
They cause ``fine structures'' in the shells of energy levels.
The peak positions, discussed above, are little affected by these
fine structures when $\lambda_1$ and $\lambda_2$ are
sufficiently small.\cite{me3}

\begin{figure}[h]
\vspace*{-0.4cm}

\begin{center}
\epsfig{file=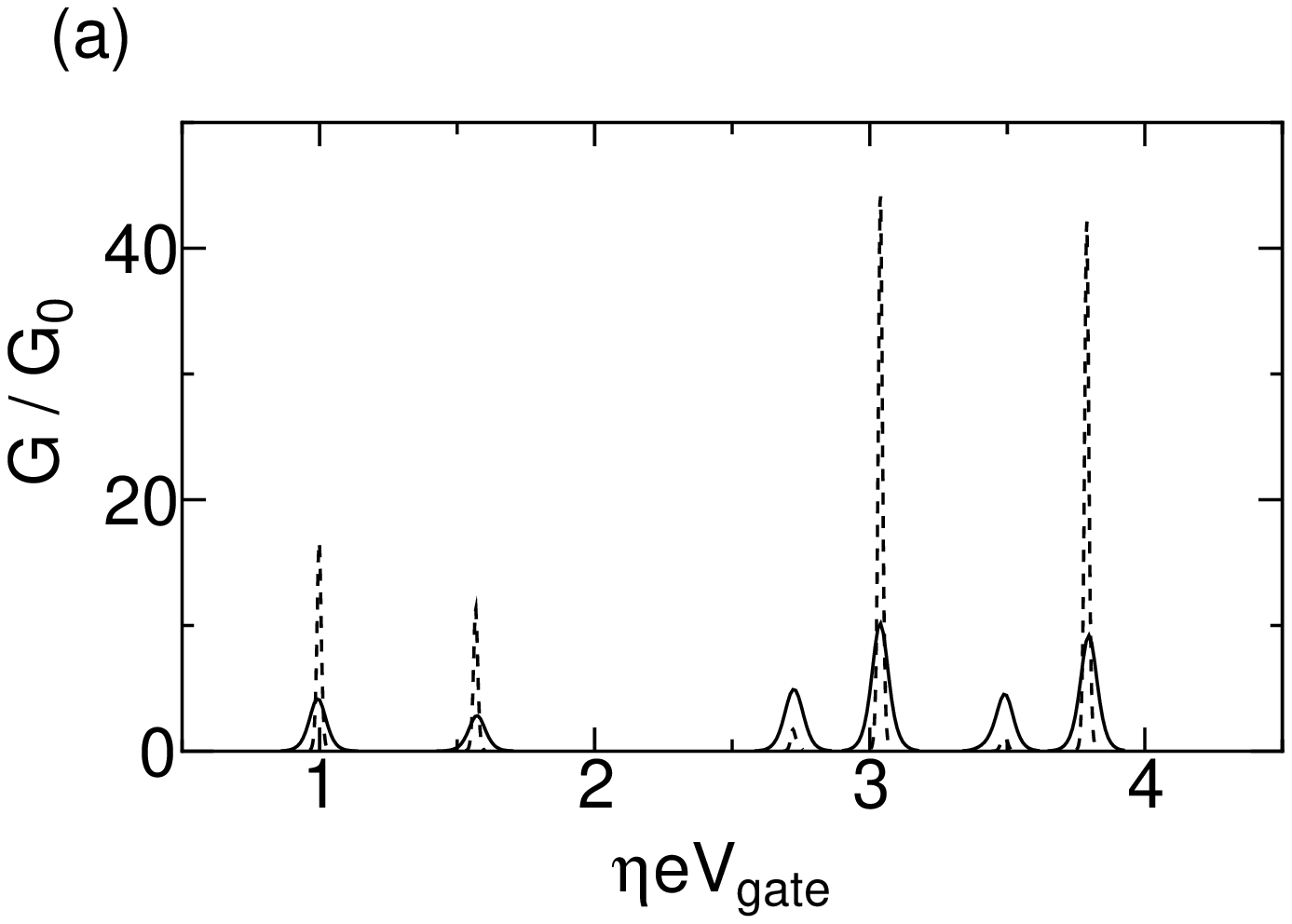,width=7.7cm}
\vspace{-0.7cm} \\
\epsfig{file=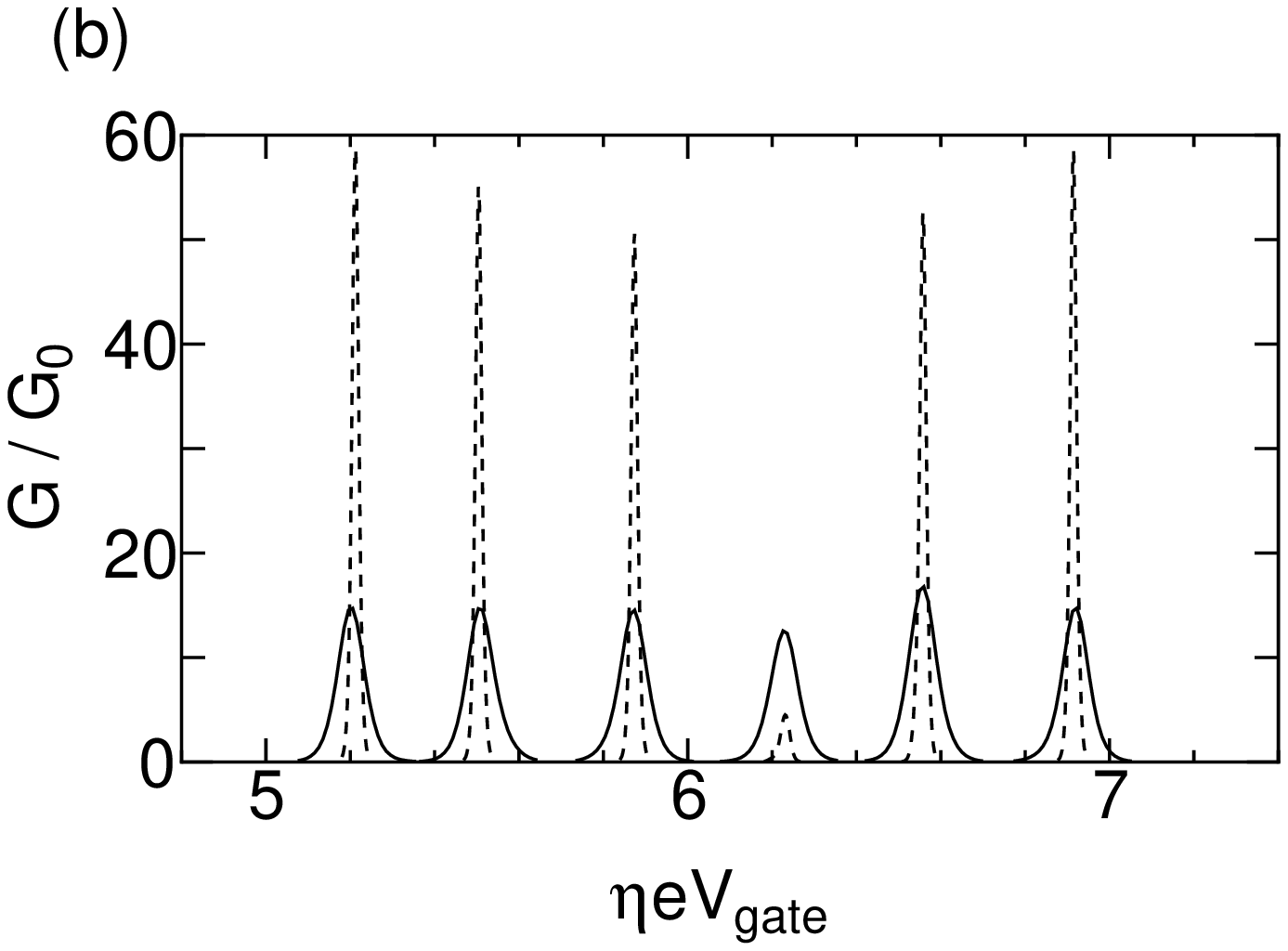,width=7.7cm}
\vspace*{-0.3cm}\\
\end{center}

\caption{
(a) The first to sixth peaks and (b) seventh to 12th peaks of the
Coulomb oscillation, in the absence of a magnetic field.
The temperature is $k_{\rm B}T/\hbar\omega_0=0.005$ (dotted lines)
and $0.02$ (solid lines).
In (a), a small anisotropy is taken into account ($\lambda_1=0.0125$).
The transmission amplitude, $V_l^{\rm L/R}$, for the lower level
in the second shell (for the first shell) is
a quarter (half) of $V_l^{\rm L/R}$ for the upper level
in the second shell. In (b), a small anharmonicity is assumed
($\lambda_2=0.025$).
$V_l^{\rm L/R}$ is constant for all the states $l$ in the dot.
The unit of $\eta eV_{\rm gate}$
($\eta=C_{\rm gate}/C_{\rm total}$) is $\hbar\omega_0$.
$G_0=(e^2/\hbar \omega_0)
\gamma^{\rm L}\gamma^{\rm R}/
(\gamma^{\rm L}+\gamma^{\rm R})$, 
where $\gamma^{\rm L/R}=(2\pi/\hbar)D^{\rm L/R}|V_l^{\rm L/R}|^2$
($l$=upper level in the second shell in (a)). 
[The strength of the Coulomb interaction is
$e^2/\varepsilon l_0=\hbar\omega_0/2$.]
\label{fig:fig4}}
\end{figure}

To explain the $T$ dependence of the third and fifth peaks, we make
two assumptions. (i) A small anisotropy of the dot ($\lambda_1=0.0125$)
splits two degenerate levels in the second shell
by $\Delta \varepsilon$ ($\approx 1.2$ K).
(ii) The transmission amplitude, $V_{{\rm L/R},l}$ in eq.\ (3.1),
for the lower level in the second shell is one-quarter of that
for the upper level.
In Fig.~\ref{fig:fig4}(a), we present the first to sixth peaks of the
linear conductance through a quantum dot as a function of gate
voltage. The third and fifth peaks are smaller in height
at $T=0.005\hbar\omega_0/k_{\rm B}$ ($\approx$ 0.3 K, dotted lines)
than at $T=0.02\hbar\omega_0/k_{\rm B}$ ($\approx$ 1 K, solid lines).

The reason for this is as follows.
When $k_{\rm B}T \ll \Delta \varepsilon$, the
third electron is almost always transported through the lower level in the
second shell, which has a smaller transmission probability to the leads
than the upper level. With increasing temperature, the upper level
contributes more to the third peak, and consequently
the third peak becomes larger. The situation is the same for the fifth
peak.\cite{me3}

When wavefunctions $\Psi_{N,i}$ are written by a single Slater
determinant, we can derive the analytical expression for the conductance.
When a current flows through a single level in the dot ($l$), the
conductance is given by
\begin{equation}
G = \frac{e^2}{4k_{\rm B} T}
\frac{\tilde{\gamma}_l}{\cosh^{2}(\delta/2k_{\rm B} T)},
\label{eq:G1}
\end{equation}
with
\begin{equation}
\tilde{\gamma}_l=\frac{\gamma_l^{\rm L} \gamma_l^{\rm R}}
                      {\gamma_l^{\rm L}+\gamma_l^{\rm R}},
\end{equation}
where $\gamma_l^{\rm L/R}=(2\pi/\hbar) D^{\rm L/R} |V_{{\rm L/R},l}|^2$ and
$\delta=E_{\rm gate}-\mu_{N,1}$.\cite{Beenakker}
In the case of the third peak,
two levels $(n,m)=(0,\pm 1)$ take part in the current.
The conductance is expressed by
\begin{eqnarray}
G = \frac{e^2}{k_{\rm B} T} \frac{2}{1+2e^{\delta/k_{\rm B} T}
+2e^{(\delta-\Delta \varepsilon)/k_{\rm B} T}}  \nonumber \\
\times \left(
\frac{\tilde{\gamma}_{(0,1)}}{1+e^{-\delta/k_{\rm B} T}}+
\frac{\tilde{\gamma}_{(0,-1)}}{1+e^{-(\delta-\Delta \varepsilon)/k_{\rm B} T}}
\right).
\label{eq:G2}
\end{eqnarray}
Here, the spin degeneracy of each level has been counted.\cite{com2}
The $T$ dependence of the conductance is characterized by two parameters,
$\Delta \varepsilon/k_{\rm B} T$ and 
$\tilde{\gamma}_{(0,-1)}/\tilde{\gamma}_{(0,1)}$.

A possible mechanism for the $T$ dependence of the 10th peak (also the
17th peak) is the ``spin blockade.''\cite{spinblock,Tanaka}
Figure \ref{fig:fig4}(b) shows the 7th to 12th peaks of the
linear conductance. A small anharmonicity is assumed
($\lambda_2=0.025$).\cite{me3} The 10th peak shows the anomalous $T$
dependence for the following reason. The ground state for $N=9$ has
the total spin of $S=3/2$ whereas the ground state for $N=10$ has $S=0$.
The 10th peak is very small at low temperatures (dotted line)
since the addition of the 10th electron on the dot is forbidden by the
spin selection rule.
With an increase in temperature, electrons are transported more smoothly
through transition-allowed excited states, and as a result, the height of
the peak grows (solid line).

\section{Transport Properties (2) Cotunneling Processes}

In the Coulomb blockade region, the current is suppressed exponentially
with decreasing temperature, as shown in eqs.\ (\ref{eq:G1}) and
(\ref{eq:G2}). In this region, however, the higher-order tunneling
processes make a significant contribution to the
transport of electrons. These processes involve the
simultaneous tunneling of more than one electron and is thus
called cotunneling.\cite{Yuli1}
There are two kinds of cotunneling processes, elastic and
inelastic processes. The latter process is accompanied by a
change of the dot state whereas the former process is not.

\begin{figure}[t]

\noindent
{\large \sf (a)} \vspace*{-.3cm} \\ \hspace*{1.7cm}
\epsfig{file=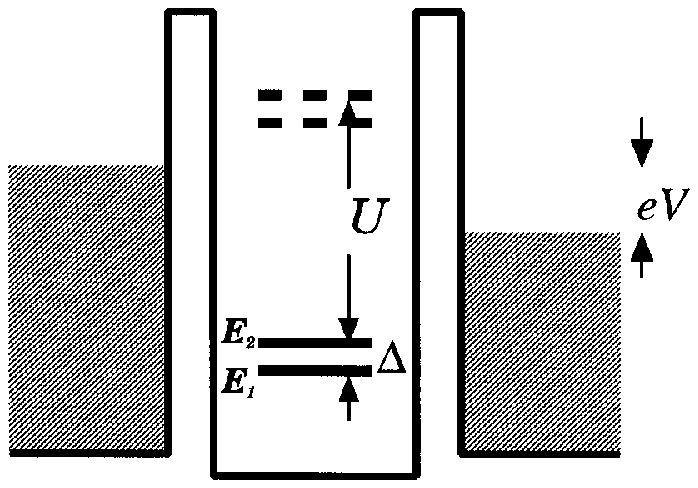,width=5.5cm}

\vspace{0.5cm}

\noindent
{\large \sf (b)} \vspace*{-.8cm} \\ \hspace*{-.8cm}
\epsfig{file=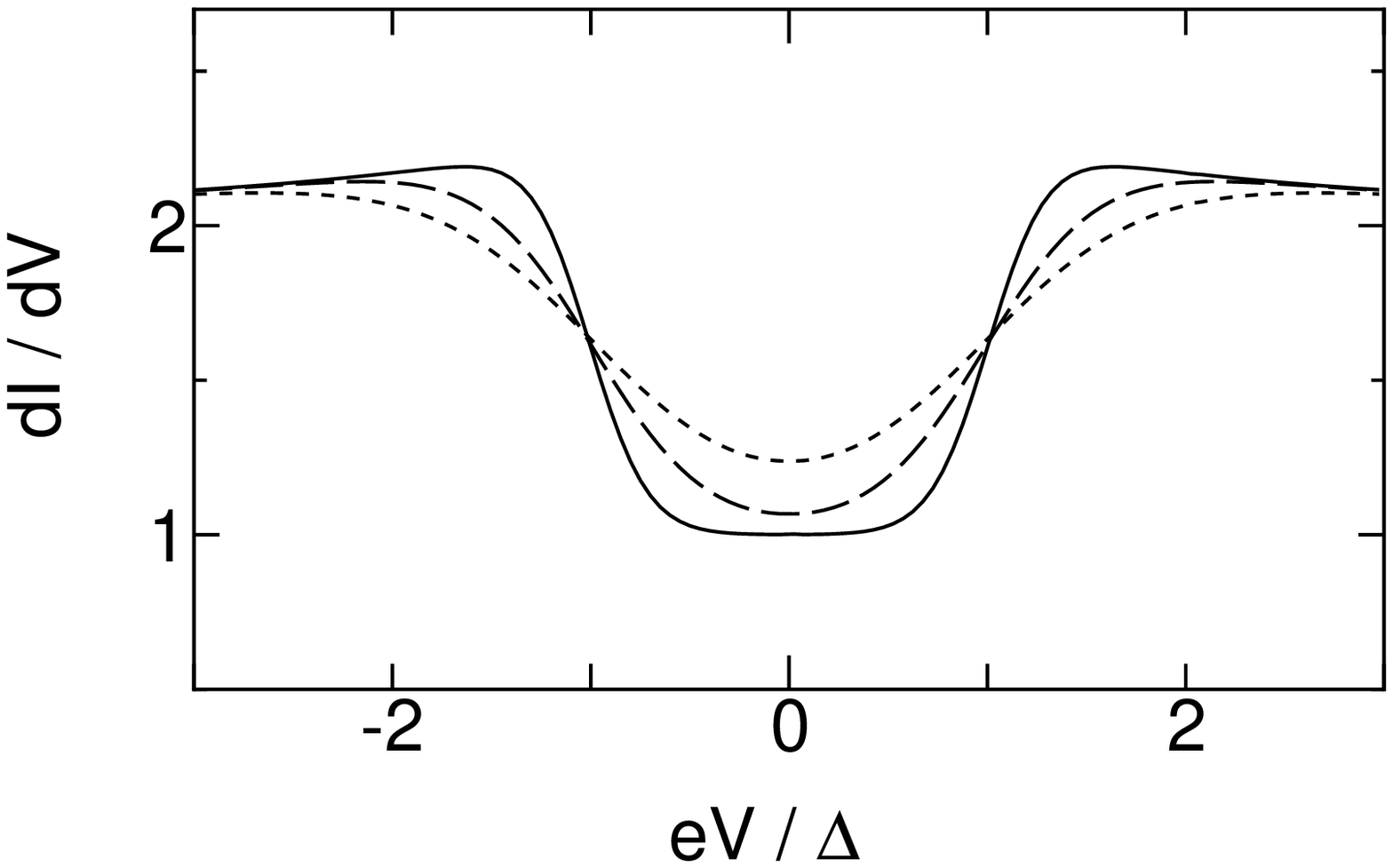,width=9cm}
\caption{
(a) A quantum dot with two levels, $E_1$ and $E_2$, and one electron
in the Coulomb blockade region. $\Delta=E_2-E_1$.
The charging energy is $U$.
Under bias voltages $V$, the cotunneling current flows.
(b) $dI/dV$ curves for the cotunneling current at $k_{\rm B}T/\Delta=$
0.1 (solid line), 0.2 (broken line) and 0.3 (dotted line).
The unit of $dI/dV$ is 
$(e^2/h) (\hbar\gamma^{\rm L}_1 \hbar\gamma^{\rm R}_1)/E_1^2$
where $\hbar\gamma^{\rm L/R}_1=2\pi D^{\rm L/R}|V_1^{\rm L/R}|^2$.
We assume that $V_1^{\rm L/R}=V_2^{\rm L/R}$, $E_1+U \gg -E_1 \gg \Delta,
e|V|$, and that $|V_1^{\rm L}/V_1^{\rm R}|^2=5$.
\label{fig:fig5}}
\end{figure}

To illustrate the cotunneling processes, let us consider a
simplified situation: one electron occupies one of
two energy levels, $E_1$ or $E_2$, in a quantum dot
(Fig.\ \ref{fig:fig5}(a)).
The energy difference between the levels is denoted by
$\Delta=E_2-E_1$. The bias voltage $V$ is applied between
two leads, $eV=\mu_{\rm L}-\mu_{\rm R}$,
where $\mu_{\alpha}$ is the Fermi level in lead $\alpha$.
The electron spin is neglected.
In the Coulomb blockade region for one electron in the dot,
$k_{\rm B}T, \hbar\gamma_i, eV
\ll -E_i,E_i+U$, where $U$ is the charging energy.
The transition probability from the dot state $i$ to $j$ with
an electron tunneling from lead $\alpha$ to $\beta$ is
\begin{eqnarray}
\Gamma^{\alpha \rightarrow \beta}_{i \rightarrow j}
& = &
\frac{2\pi}{\hbar}
\sum_{k,k'}
\left| T^{\alpha \rightarrow \beta}_{i \rightarrow j} \right|^2
\delta(\varepsilon_{\alpha, k}+E_i-\varepsilon_{\beta, k'}-
E_j)   \nonumber \\
& & \times  f(\varepsilon_{\alpha, k}-\mu_{\alpha})
\left[1-f(\varepsilon_{\beta, k'}-\mu_{\beta}) \right] \nonumber \\
& = &
\frac{2\pi}{\hbar}
\left| T^{\alpha \rightarrow \beta}_{i \rightarrow j} \right|^2
D^{\alpha} D^{\beta} F(E_i-E_j+\mu_{\alpha}-\mu_{\beta}),
\end{eqnarray}
where
\[ 
F(\varepsilon)=\frac{\varepsilon}{1-e^{-\varepsilon/k_{\rm B}T}}.
\]
$T^{\alpha \rightarrow \beta}_{i \rightarrow j}$ is
the component of the T-matrix
\begin{eqnarray}
T^{\alpha \rightarrow \beta}_{1 \rightarrow 1}
& \approx &
-\frac{V_{\beta,2}^* V_{\alpha,2}}{E_2+U}-
\frac{V_{\beta,1}^* V_{\alpha,1}}{E_1} 
\label{eq:T11} \\
T^{\alpha \rightarrow \beta}_{1 \rightarrow 2}
& \approx &
-V_{\beta,1}^* V_{\alpha,2}
\left(\frac{1}{E_2+U}-\frac{1}{E_1} \right),
\label{eq:T12}
\end{eqnarray}
etc., to the second order of
${\cal H}_{\rm T}$.\cite{Funabas1,Funabas2}
The processes of $i=j$ are elastic whereas the
processes of $i \ne j$ are inelastic. In the
latter, the conduction electron loses or gains
an energy of $\Delta$.
Two terms on the right side of eqs.\ 
(\ref{eq:T11}) and (\ref{eq:T12})
stem from two virtual states in which
the dot is doubly occupied or unoccupied.
It should be noted that the interference
between the two virtual processes always
enhances the inelastic cotunneling whereas
the interference suppresses (enhances)
the elastic cotunneling when
$V_{{\rm R},1}^* V_{{\rm L},1}$ and
$V_{{\rm R},2}^* V_{{\rm L},2}$
are in phase (out of phase).\cite{Funabas2}

The cotunneling current is expressed as
\begin{equation}
I=e\sum_{i,j=1,2} \rho_i \left(
\Gamma^{{\rm L} \rightarrow {\rm R}}_{i \rightarrow j}-
\Gamma^{{\rm R} \rightarrow {\rm L}}_{i \rightarrow j}
\right),
\end{equation}
where $\rho_i$ is the probability of dot state $i$
($\rho_1+\rho_2=1$). They are determined by the
stationary condition\cite{Yuli2}
\begin{equation}
\frac{d}{dt} \rho_1=
-\rho_1 \sum_{\alpha,\beta={\rm L,R}}
\Gamma^{\alpha \rightarrow \beta}_{1 \rightarrow 2}
+{\rho_2} \sum_{\alpha,\beta={\rm L,R}}
\Gamma^{\alpha \rightarrow \beta}_{2 \rightarrow 1}=0.
\end{equation}
Figure \ref{fig:fig5}(b) shows the calculated results of
$dI/dV$ as a function of $V$.
At $k_{\rm B}T \ll \Delta$, the $dI/dV$ curve appears to be
a step function. When $|eV|<\Delta$, only the elastic
process is possible because the conduction electron cannot
lose the energy $\Delta$ (the dot state is almost always $1$
at $k_{\rm B}T \ll \Delta$). When $|eV|>\Delta$,
both the elastic and inelastic processes
contribute to the current, which increases the conductance.\cite{com3}
The onset of the inelastic process is smeared with increasing
$T$. These properties of the cotunneling conductance have
been observed experimentally.\cite{Silvano}

Next, let us consider the spin of electrons but disregard
the upper level $E_2$. Some cotunneling processes
give rise to spin-flips in the dot, {\it e.g.}, immediately after an
up-spin electron in the dot goes to lead $\beta$,
a down-spin electron enters the dot from lead $\alpha$.
The spin-flip processes can be expressed,
to the second order of ${\cal H}_{\rm T}$,
by the effective Hamiltonian
\begin{eqnarray}
{\cal H}_{\rm eff}=\sum_{\alpha,\beta={\rm L/R}}\sum_{k,k'}
J_{\beta\alpha}
\Bigl[
 \hat{S}_{-} c_{\beta k' \uparrow}^{\dagger} c_{\alpha k \downarrow}
+\hat{S}_{+} c_{\beta k' \downarrow}^{\dagger} c_{\alpha k \uparrow}
\nonumber \\
+\hat{S}_{z} (c_{\beta k' \uparrow}^{\dagger} c_{\alpha k \uparrow}
        -c_{\beta k' \downarrow}^{\dagger} c_{\alpha k \downarrow}),
\Bigr]
\label{eq:sdH}
\end{eqnarray}
with
\begin{equation}
J_{\beta\alpha}=V_{\beta,1}^* V_{\alpha,1}
\left(\frac{1}{E_1+U}-\frac{1}{E_1} \right),
\end{equation}
where $\hat{S}$ is the spin operator for the dot state;
$\hat{S}_{-}=d_{1,\downarrow}^{\dagger} d_{1,\uparrow}$,
$\hat{S}_{+}=d_{1,\uparrow}^{\dagger} d_{1,\downarrow}$,
$\hat{S}_{z}=(d_{1,\uparrow}^{\dagger} d_{1,\uparrow}-
              d_{1,\downarrow}^{\dagger} d_{1,\downarrow})/2$.
This is the Kondo Hamiltonian which was originally used to describe the
exchange coupling between a spin of magnetic impurities
and conduction electrons in the host metal.\cite{classics,classics2}
The Kondo effect in quantum dots has been observed
experimentally.\cite{Kondo1,Kondo2,Kondo3,Kondo4,Kondo5}
When the number of electrons in a dot is odd, the
localized spin $S=1/2$ in the dot is coupled to the Fermi sea
in the leads, which results in the formation of the Kondo resonance
at the Fermi level. The cotunneling conductance is
markedly enhanced through this resonant level
to a value of $2e^2/h$, at low temperatures of $T \ll T_{\rm K}$.
$T_{\rm K}$ is the Kondo temperature which is given by
\begin{equation}
k_{\rm B}T_{\rm K} \approx \sqrt{|E_1(E_1+U)|} \exp \left[-1/2D
\sqrt{J_{\rm LL}^2+J_{\rm RR}^2} \right],
\end{equation}
where $D$ is the density of states in the leads.\cite{classics,classics2}
This effect is usually irrelevant for even numbers of electrons
in the dot.

Recently a large Kondo effect has been observed for an
even number of electrons when spin-triplet and singlet
states are degenerate.\cite{Sasaki}
The energy difference between these states can be
tuned by the magnetic field, as discussed in \S 2.
This is a situation unique to quantum dots and cannot be
realized in traditional Kondo systems of magnetic impurities in metals.
We have theoretically examined this situation.
We have assumed that two electrons occupy
levels $E_1$ or $E_2$ in Fig.\ \ref{fig:fig5}(a), making spin-triplet
or singlet states.
By extending the spin-flip Hamiltonian (\ref{eq:sdH})
and analyzing it,
we have revealed an enhancement of $T_{\rm K}$ due to
the competition between spin-triplet and singlet
states.\cite{meYuli}

\section{Transport through Dot Molecules}

Artificial molecules in which two quantum dots are connected
in series have been fabricated on a
semiconductor.\cite{Oosterkamp,Blick}
The strength of the tunneling between the quantum dots can be
controlled electrostatically.
We examine the Coulomb oscillation through an
artificial molecule in which two equivalent quantum dots
(dot 1 and dot 2) are connected in series.
A common gate is attached to the dots.

We consider the simplest case in which one level in each
quantum dot is relevant for the transport.
The tunneling Hamiltonian describes the coupling between dot 1 and
the left lead ($V_1^{\rm L}$) and between dot 2 and the right lead
($V_2^{\rm R}$). The couplings are assumed to be sufficiently small,
compared with the tunneling between the dots, $t (<0)$,
which allows the formation of molecular orbitals between the
dots.\cite{Kawamura}
The many-body states in the molecule are obtained exactly, taking
the intradot Coulomb interaction ($U_0$) and
interdot Coulomb interaction ($U_1$) into consideration.\cite{me4}

In the ground state for $N=1$ in the molecule, an electron occupies
the bonding orbital,
$\psi_{\rm B}=(|1\rangle+|2\rangle)/\sqrt{2}$, where $|n\rangle$
is the orbital in dot $n$.
The ground state for $N=2$ is a spin singlet and given by
\begin{eqnarray}
\Psi_{2,1} & = & |1\uparrow 1\downarrow \rangle +
                     |2\uparrow 2\downarrow \rangle   \nonumber \\
   & + & \frac{1}{4} \left( \tilde{U}+\sqrt{\tilde{U}^2+16} \right)
        \left( |1\uparrow 2\downarrow \rangle +
               |2\uparrow 1\downarrow \rangle \right),
\label{eq:dbleQD}
\end{eqnarray}
apart from a normalization constant.
The parameter $\tilde{U}=(U_0-U_1)/|t|$
characterizes the strength of the correlation effect.
When $\tilde{U}^<_{\sim} 1$, the correlation effect is weak. In the ground
state, two electrons occupy the bonding orbital (eq.\ (\ref{eq:dbleQD})
$\approx |\psi_{\rm B} \uparrow \psi_{\rm B} \downarrow \rangle$).
When $\tilde{U}\gg 1$, the strong correlation brings about
the Heitler-London wavefunction of the ground state,
in which each electron is localized in a dot 
(eq.\ (\ref{eq:dbleQD}) $\approx |1\uparrow 2\downarrow \rangle +
|2\uparrow 1\downarrow \rangle$).\cite{H-L}

\begin{figure}[t]
\vspace*{0.3cm}

\noindent
{\large \sf (a)} \vspace{-.5cm} \\ \hspace{-1cm}
\epsfig{file=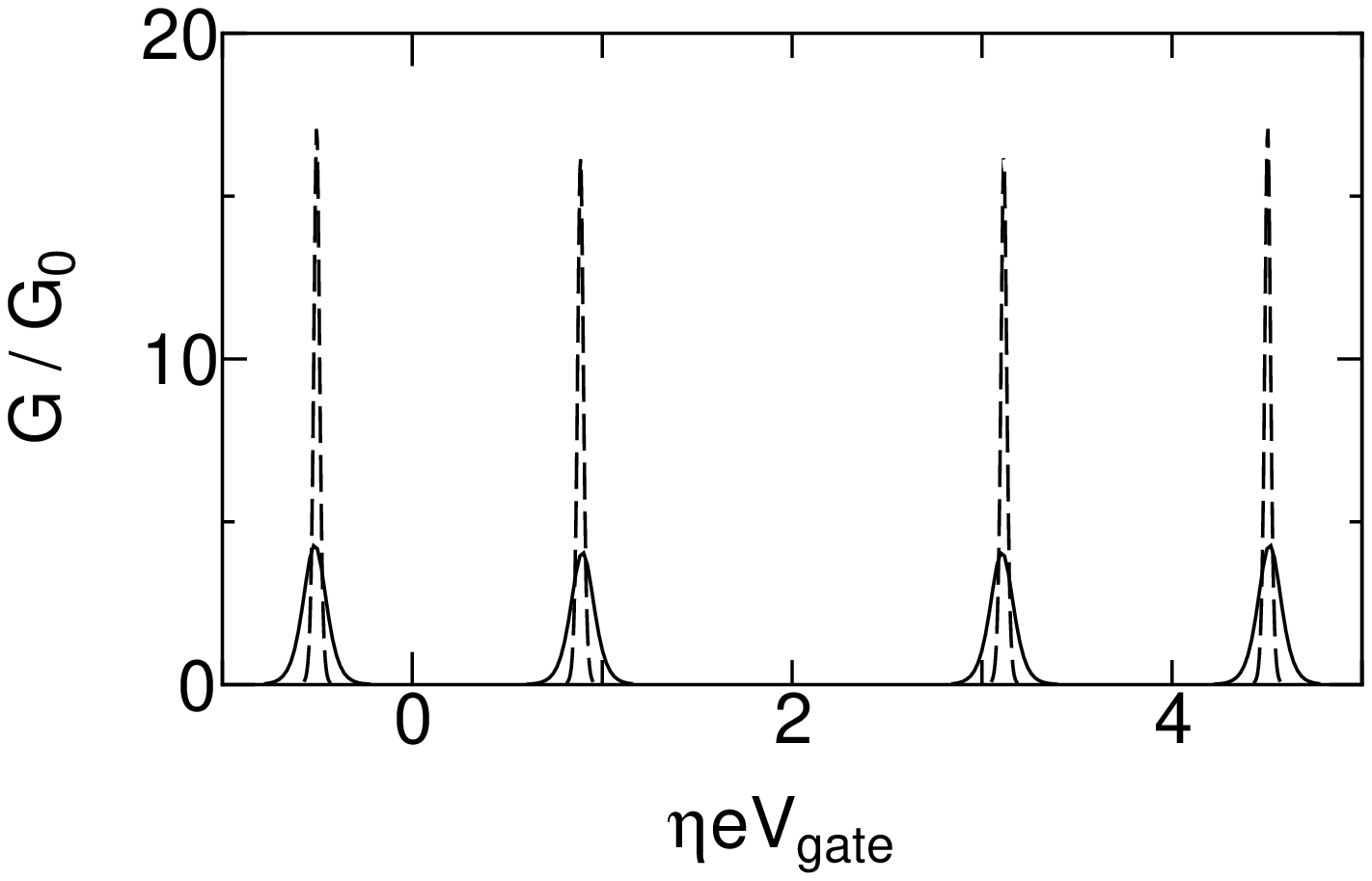,width=7.7cm}

\vspace{-.5cm}

\noindent
{\large \sf (b)} \vspace{-.5cm} \\ \hspace{-1cm}
\epsfig{file=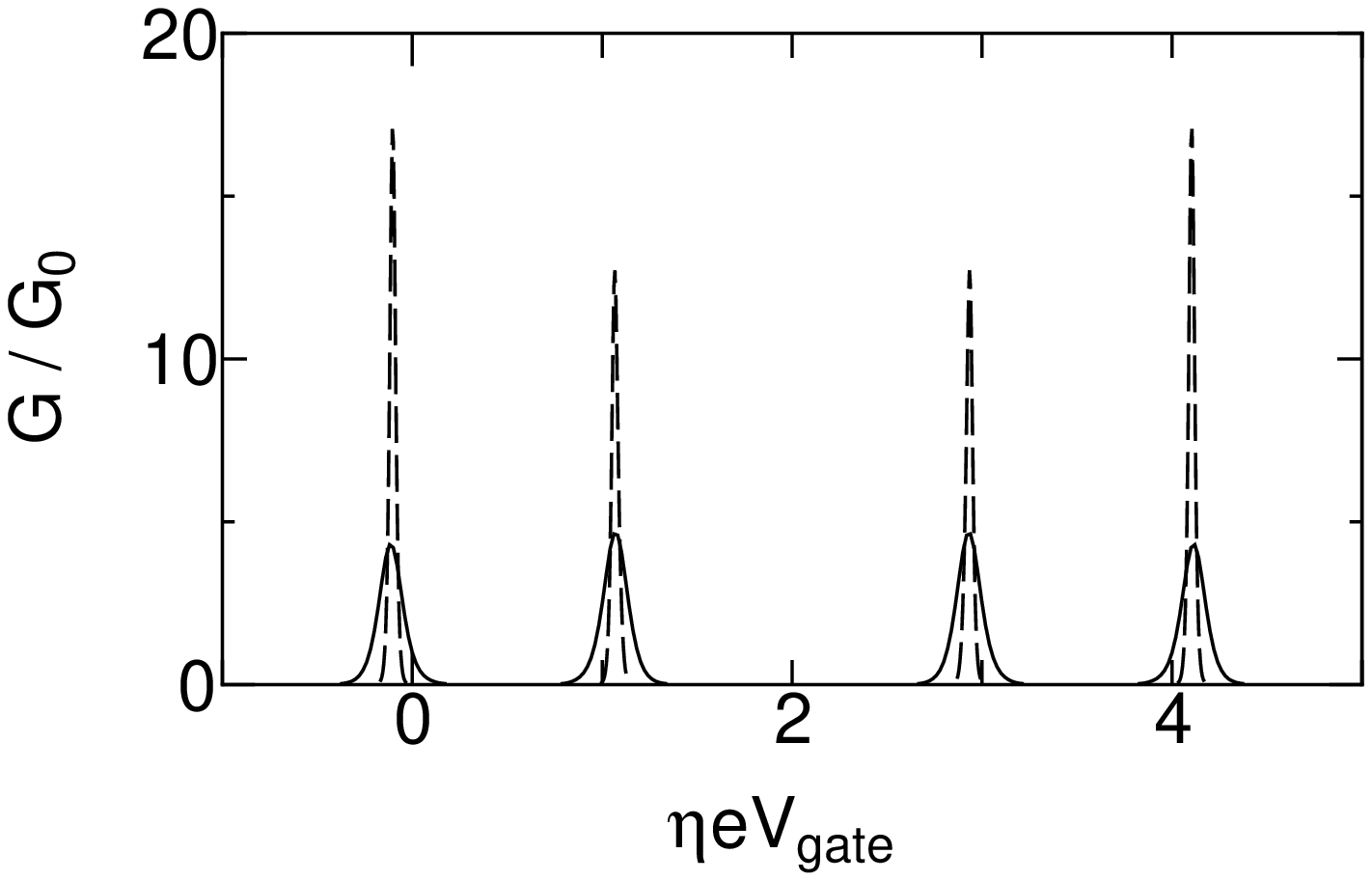,width=7.7cm}

\caption{
The gate voltage dependence of the linear conductance
$G$ through two coupled dots. $U_0=2U_1$. The temperature is
$k_{\rm B}T/U_1=0.01$ (broken lines), $0.04$ (solid lines).
(a) $|t|=0.5U_1$ ($\tilde{U}=2$) and (b) $|t|=0.1U_1$ ($\tilde{U}=10$).
The unit of $\eta eV_{\rm gate}$ 
($\eta=C_{\rm gate}/C_{\rm total}$) is $U_1$.
$G_0=(e^2/U_1)\gamma^{\rm L}\gamma^{\rm R}/(\gamma^{\rm L}+\gamma^{\rm R})$,
where $\displaystyle \gamma^{\rm L}=(2\pi/\hbar)D^{\rm L}
\left|V_1^{\rm L} \right|^2$ and
$\displaystyle \gamma^{\rm R}=(2\pi/\hbar)D^{\rm R}
\left|V_2^{\rm R} \right|^2$.
\label{fig:fig6}}
\end{figure}

In Fig.~\ref{fig:fig6}, we present the gate voltage dependence of the
linear conductance $G$ when $U_0=2U_1$ and
(a) $|t|=0.5U_1$ ($\tilde{U}=2$), (b) $|t|=0.1U_1$ ($\tilde{U}=10$).
In case (b), the second and third peaks are suppressed
at low temperatures ($k_{\rm B}T/U_1=0.01$, broken lines).
This is due to the strong correlation effect, as explained in the following.
In case (a) of weak correlation, the second electron enters dot 1 from
the left lead and occupies the bonding orbital $\psi_{\rm B}$. Thus the
tunneling probability is proportional to
$|\langle 1 | \psi_{\rm B} \rangle|^2=1/2$.
In case (b) of strong correlation, the first electron (say, with spin
$\downarrow$) has to be in dot 2 for the second electron ($\uparrow$)
to enter the dot 1. This probability is $1/2$. Then the state is
$|1 \uparrow 2 \downarrow \rangle$, which has the probability of $1/2$
in the ground state $\Psi_{2,1}$. Hence the total probability is
$(1/2)\times(1/2)=1/4$.
The strong correlation, therefore, reduces the height of the second peak
(also the third peak) by one-half.\cite{me4}

In the case of strong correlation, electron spins localized in the dots
are coupled antiferromagnetically to each other.
If the localized spins are regarded as ``qubits'' for the
quantum computer, the spin-singlet state $\Psi_{2,1}$ is
an ``entangled'' state between them. The strength of the antiferromagnetic
coupling can be tuned by gate voltages,
magnetic or electric fields, which is applicable to the
switching of the qubit-qubit interaction.\cite{Loss1,Loss2}

\section{Conclusions}

The electronic states in quantum dots have been calculated, taking
into account the electron-electron interaction exactly.
The magnetic field dependences of the ground state and low-lying excited
states agree well with the experimental results.
The conductance peaks of the Coulomb oscillation have been examined,
using the obtained many-body states. We have proposed two mechanisms
for the anomalous $T$ dependence of some peaks.
In the Coulomb blockade region, elastic and inelastic
cotunneling processes make a significant contribution to the current.
The Kondo effect may extremely enhance the cotunneling conductance.
In the transport through coupled two dots, the correlation
effect influences the peak heights of the Coulomb oscillation.

The agreement between the calculated results and experimental ones
will enable further research using quantum dots for unsolved problems
in solid-state physics. Among them are the many-body problems: the
correlation effect on the electronic states and on the transport
properties and the Kondo effect. Under finite bias voltages, we have only
discussed the cotunneling processes to the second order of the tunneling
Hamiltonian. This is an essentially nonequilibrium problem which requires
more study both theoretically and experimentally.
The inelastic process results in the dephasing of the conduction
electrons.\cite{Funabas1,Funabas2}
It must be necessary to control the dephasing processes, including
other mechanisms, {\it e.g.}, electron-phonon interaction,
for the application of quantum dots to quantum computers.

\acknowledgements

The author thanks S.\ Tarucha, L.\ P.\ Kouwenhoven, Yu.\ V.\
Nazarov, G.\ E.\ W.\ Bauer, T.\ Aono and K.\ Kawamura
for valuable discussions.
This work was partly supported by a Grant-in-Aid
from the Ministry of Education, Science, Sports and Culture.

\pagebreak

\pagebreak

\begin{thebibliography}{99}
\bibitem{Tarucha}
S.\ Tarucha, D.\ G.\ Austing, T.\ Honda, R.\ J.\ van der Hage and
L.\ P.\ Kouwenhoven: Phys.\ Rev.\ Lett.\ {\bf 77} (1996) 3613.
\bibitem{Oosterkamp}
T.\ H.\ Oosterkamp, T.\ Fujisawa, W.\ G.\ van der Wiel, K.\ Ishibashi,
R.\ V.\ Hijman, S.\ Tarucha and L.\ P.\ Kouwenhoven:
Nature {\bf 395} (1998) 873.
\bibitem{Blick}
R.\ H.\ Blick, D.\ Pfannkuche, R.\ J.\ Haug, K.\ v.\ Klitzing
and K.\ Eberl: Phys.\ Rev.\ Lett.\ {\bf 80} (1998) 4032.
\bibitem{qd1}
G.\ W.\ Bryant: Phys.\ Rev.\ Lett.\
{\bf 59} (1987) 1140.
\bibitem{qd2}
P.\ A.\ Maksym and T.\ Chakraborty: Phys.\ Rev.\ Lett.\
{\bf 65} (1990) 108.
\bibitem{qd3}
P.\ Hawrylak: Phys.\ Rev.\ Lett.\ {\bf 71} (1993) 3347.
\bibitem{spinblock}
D.\ Weinmann, W.\ H\"{a}usler and B.\ Kramer: Phys.\ Rev.\ Lett.\
{\bf 74} (1995) 984.
\bibitem{Daniela}
D.\ Pfannkuche and S.\ E.\ Ulloa:
Phys.\ Rev.\ Lett.\ {\bf 74} (1995) 1194.
\bibitem{me1}
M.\ Eto: Jpn.\ J.\ Appl.\ Phys.\ {\bf 36} (1997) 3924.
\bibitem{me2}
M.\ Eto: Jpn.\ J.\ Appl.\ Phys.\ {\bf 38} (1999) 376.
\bibitem{Leo}
L.\ P.\ Kouwenhoven, T.\ H.\ Oosterkamp, M.\ W.\ S.\ Danoesastro,
M.\ Eto, D.\ G.\ Austing, T.\ Honda and S.\ Tarucha:
Science {\bf 278} (1997) 1788.
\bibitem{me3}
M.\ Eto: J.\ Phys.\ Soc.\ Jpn.\ {\bf 66} (1997) 2244.
\bibitem{Yuli1}
D.\ V.\ Averin and Yu.\ V.\ Nazarov:
Phys.\ Rev.\ Lett.\ {\bf 65} (1990) 2446.
\bibitem{Funabas1}
Y.\ Funabashi, K.\ Ohtsubo, M.\ Eto and K.\ Kawamura:
Solid-State Electron.\ {\bf 42} (1998) 1367.
\bibitem{Funabas2}
Y.\ Funabashi, K.\ Ohtsubo, M.\ Eto and K.\ Kawamura:
Jpn.\ J.\ Appl.\ Phys.\ {\bf 38} (1999) 388.
\bibitem{Kondo1}
D.\ Goldhaber-Gordon, H.\ Shtrikman, D.\ Mahalu, D.\ Abusch-Magder,
U.\ Meirav and M.\ A.\ Kastner: Nature {\bf 391} (1998) 156.
\bibitem{Kondo2}
D.\ Goldhaber-Gordon, J.\ G\"ores, M.\ A.\ Kastner, H.\ Shtrikman, D.\ Mahalu
and U.\ Meirav: Phys.\ Rev.\ Lett.\ {\bf 81} (1998) 5225.
\bibitem{Kondo3}
S.\ M.\ Cronenwett, T.\ H.\ Oosterkamp and L.\ P.\ Kouwenhoven:
Science {\bf 281} (1998) 540.
\bibitem{Kondo4}
F.\ Simmel, R.\ H.\ Blick, J.\ P.\ Kotthaus, W.\ Wegscheider and M.\ Bichler:
Phys.\ Rev.\ Lett.\ {\bf 83} (1999) 804.
\bibitem{Kondo5}
J.\ Schmid, J.\ Weis, K.\ Eberl and K.\ v.\ Klitzing:
Phys.\ Rev.\ Lett. {\bf 84} (2000) 5824.
\bibitem{Kawamura}
K.\ Kawamura and T.\ Aono: Jpn.\ J.\ Appl.\ Phys.\ {\bf 36} (1997)
3951.
\bibitem{Loss1}
G.\ Burkard, D.\ Loss and D.\ P.\ DiVincenzo: Phys.\ Rev.\ B
{\bf 59} (1999) 2070.
\bibitem{Loss2}
G.\ Burkard, D.\ Loss and D.\ P.\ DiVincenzo: Phys.\ Rev.\ B
{\bf 62} (2000) 2581.
\bibitem{me4}
M.\ Eto: Solid-State Electron.\ {\bf 42} (1998) 1373.
\bibitem{com1}
In the case of low magnetic fields and the Coulomb interaction strength
of $e^2/\varepsilon l_0=\hbar\omega_0$,
we consider all the configurations in which $N$ electrons occupy
15 energy levels in the first to fifth shells. For $N=4$, for example,
the dimension of the configuration space with the $z$ component of the total
spin $S_z=0$ [and total angular momentum $M=0$] is
$(_{15}C_2)^2=11025$ [$1025$].
\bibitem{Tarucha2}
S.\ Tarucha, D.\ G.\ Austing, Y.\ Tokura, W.\ G.\ van der Wiel
and L.\ P.\ Kouwenhoven: Phys.\ Rev.\ Lett.\ {\bf 84}
(2000) 2485.
\bibitem{Wagner}
M.\ Wagner, U.\ Merkt and A.\ V.\ Chaplik: Phys.\ Rev.\ B {\bf 45}
(1992) 1951.
\bibitem{MDD}
T.\ H.\ Oosterkamp, J.\ W.\ Janssen, L.\ P.\ Kouwenhoven, D.\ G.\
Austing, T.\ Honda and S.\ Tarucha: Phys.\ Rev.\ Lett.\ {\bf 82}
(1999) 2931.
\bibitem{Natori}
A.\ Natori, Y.\ Sugimoto and M.\ Fujito:
Jpn.\ J.\ Appl.\ Phys.\ {\bf 36} (1997) 3960.
\bibitem{Beenakker}
C.\ W.\ Beenakker: Phys.\ Rev.\ B {\bf 44} (1991) 1646.
\bibitem{Tanaka}
Y.\ Tanaka and H.\ Akera: Phys.\ Rev.\ B {\bf 53} (1996) 3901.
\bibitem{com0}
This treatment is valid at the current peaks
when $k_{\rm B}T \gg \hbar \Gamma_{N+1,j; N,i}^{\alpha}$.
\bibitem{Leo2}
R.\ J.\ van der Hage: Ph.\ D thesis, Delft University of Technology (1996).
\bibitem{com2}
In eq.\ (\ref{eq:G1}), the electron spin is neglected. If the
spin degeneracy of level $l$ is counted, this equation is replaced by
\[
G = \frac{e^2}{k_{\rm B} T} \tilde{\gamma}_{l}
\frac{1}{1+2e^{\delta/k_{\rm B} T}}
\frac{2}{1+e^{-\delta/k_{\rm B} T}}.
\]
\bibitem{Yuli2}
Yu.\ V.\ Nazarov: Physica B {\bf 189} (1993) 57.
\bibitem{com3}
Around the onset of the inelastic process, an anomalous hump of $dI/dV$
is evident on the step function at low temperatures.
This is attributable to the change of the probability of dot state $2$,
$\rho_2$, as a function of $V$. Around $e|V| \sim \Delta$, $\rho_2$
increases from zero to a finite value which is determined by eq.\ (4.5).
This results in the extra increase of the current, eq.\ (4.4), in addition to
the increase of $\Gamma^{{\rm L} \rightarrow {\rm R}}_{1 \rightarrow 2}$
owing to the inelastic process. The range of the bias voltage for this
hump is determined by $k_{\rm B}T$.
\bibitem{Silvano}
S.\ De Franceschi, J.\ M.\ Elzerman, W.\ G.\ van der Wiel,
L.\ P.\ Kouwenhoven and S.\ Tarucha: cond-mat/0007448,
Phys.\ Rev.\ Lett.\ {\bf 86}, No.\ 5 (2001), in press.
\bibitem{classics}
A.\ C.\ Hewson: {\it The Kondo Problem to Heavy Fermions}
(Cambridge, Cambridge, 1993).
\bibitem{classics2}
K.\ Yosida: {\it Theory of Magnetism} (Springer, New York, 1996).
\bibitem{Sasaki}
S.\ Sasaki, S.\ De Franceschi, J.\ M.\ Elzerman, W.\ G.\ van der Wiel,
M.\ Eto, S.\ Tarucha and L.\ P.\ Kouwenhoven:
Nature {\bf 405} (2000) 764.
\bibitem{meYuli}
M.\ Eto and Yu.\ V.\ Nazarov:
Phys.\ Rev.\ Lett.\ {\bf 85} (2000) 1306.
M.\ Eto and Yu.\ V.\ Nazarov: cond-mat/0101152.
\bibitem{H-L}
W.\ Heitler and F.\ London: Z.\ Phys.\ {\bf 44} (1927) 455.
\end{thebibliography}
\end{document}